\documentclass[conference,10pt]{IEEEtran}
\IEEEoverridecommandlockouts
\usepackage{booktabs} 
\usepackage{url} 
\usepackage{graphicx} 
\usepackage{amsmath,amssymb,amsfonts} 
\usepackage{xcolor} 
\usepackage[font=small,labelfont=bf]{caption} 
\usepackage{textcomp} 
\usepackage{fancyhdr}

\usepackage{dblfloatfix}
\usepackage{cite}
\usepackage{soul} 
\usepackage{enumitem} 
\usepackage{lipsum} 
\usepackage[hang,flushmargin]{footmisc}
\usepackage{etoolbox}
\usepackage{float}
\usepackage{multirow}
\usepackage{hhline} 
\usepackage{multicol}
\usepackage{amssymb}
\newlist{inlinelist}{enumerate*}{1}
\setlist[inlinelist]{label=(\arabic*)}

\newcommand{\changed}[1]{\textcolor{black}{#1}}

\usepackage{algorithmicx}
\usepackage{algorithm}
\usepackage{algpseudocode}
\usepackage{wrapfig}

\algnewcommand\algorithmicinput{\textbf{Input:}}
\algnewcommand\Input{\item[\algorithmicinput]}

\algnewcommand\algorithmicconst{\textbf{Constraints:}}
\algnewcommand\Const{\item[\algorithmicconst]}

\algnewcommand\algorithmicoutput{\textbf{Output:}}
\algnewcommand\Output{\item[\algorithmicoutput]}

\algnewcommand{\algorithmicgoto}{\textbf{go to}}%
\algnewcommand{\Goto}[1]{\algorithmicgoto~\ref{#1}}%

\algrenewcommand\algorithmicindent{0.5em}

\renewcommand{\thefigure}{\arabic{figure}}

\usepackage{tikz}

\usepackage{array}
\newcolumntype{L}[1]{>{\raggedright\let\newline\\\arraybackslash\hspace{0pt}}m{#1}}
\newcolumntype{C}[1]{>{\centering\let\newline\\\arraybackslash\hspace{0pt}}m{#1}}
\newcolumntype{R}[1]{>{\raggedleft\let\newline\\\arraybackslash\hspace{0pt}}m{#1}}
\newcolumntype{M}[1]{>{\centering\arraybackslash}m{#1}}
\newcolumntype{O}[1]{>{\raggedleft\arraybackslash}m{#1}}
\usepackage{makecell}
\usepackage{diagbox}

\usepackage[hidelinks, bookmarks=false]{hyperref}

\def\BibTeX{{\rm B\kern-.05em{\sc i\kern-.025em b}\kern-.08em
    T\kern-.1667em\lower.7ex\hbox{E}\kern-.125emX}}
\begin{document}

\title{Xel-FPGAs: An End-to-End Automated Exploration Framework for Approximate Accelerators in FPGA-Based Systems}
\newcommand{\vm}[1]{\textcolor{red}{#1}}

\author{
    \IEEEauthorblockN{
    Bharath Srinivas Prabakaran\IEEEauthorrefmark{1}\textsuperscript{,}\IEEEauthorrefmark{4}\thanks{\IEEEauthorrefmark{4}These two authors have contributed to this work equally.}, Vojtech Mrazek\IEEEauthorrefmark{2}\textsuperscript{,}\IEEEauthorrefmark{4}, Zdenek Vasicek \IEEEauthorrefmark{2}, Lukas Sekanina\IEEEauthorrefmark{2}, Muhammad Shafique\IEEEauthorrefmark{3}}
    \IEEEauthorblockA{\IEEEauthorrefmark{1}Institute of Computer Engineering, Technische Universit{\"a}t Wien (TU Wien), Austria
    \\bharath.prabakaran@tuwien.ac.at}
    \IEEEauthorblockA{\IEEEauthorrefmark{2}Faculty of Information Technology, IT4Innovations Centre of Excellence, Brno University of Technology, Czech Republic
    \\\{mrazek, vasicek, sekanina\}@fit.vutbr.cz}
    \IEEEauthorblockA{\IEEEauthorrefmark{3}eBrain Lab, Division of Engineering, New York University Abu Dhabi (NYUAD), United Arab Emirates (UAE)
    \\muhammad.shafique@nyu.edu}
}



\fancypagestyle{firstpage}{
  \fancyhf{}
  \renewcommand{\headrulewidth}{0.4pt}
  \fancyhead[C]{To appear at the 42nd International Conference on Computer-Aided Design (ICCAD), November 2023, San Francisco, CA.}
}

\maketitle
\thispagestyle{firstpage}
\pagestyle{plain}

\begin{abstract}
Generation and exploration of approximate circuits and accelerators has been a prominent research domain to achieve energy-efficiency and/or performance improvements.
This research has predominantly focused on ASICs, while not achieving similar gains when deployed for FPGA-based accelerator systems, due to the inherent architectural differences between the two.
In this work, we propose a novel framework, Xel-FPGAs, which leverages statistical or machine learning models to effectively explore the architecture-space of state-of-the-art ASIC-based approximate circuits to cater them for FPGA-based systems given a simple RTL description of the target application.
We have also evaluated the scalability of our framework on a multi-stage application using a hierarchical search strategy.
The Xel-FPGAs framework is capable of reducing the exploration time by up to $95\%$, when compared to the default synthesis, place, and route approaches, while identifying an improved set of Pareto-optimal designs for a given application, when compared to the state-of-the-art.
The complete framework is open-source and available online at \textcolor{blue}{\url{https://github.com/ehw-fit/xel-fpgas}}.
\end{abstract}



\section{Introduction}
\label{sec:Intro}
Field Programmable Gate Arrays (FPGAs) act as a lucrative computing platform in various applications, especially ones that need to be updated and/or modified on-the-go, due to their ability of partial run-time reconfiguration, \textit{i.e.}, field-programmability.
Since their introduction in $1984$~\cite{trimberger2018three}, FPGAs have gained a lot of popularity due to their decreased time-to-market and lower prototype costs, when compared to Application Specific Integrated Circuits (ASICs).
Besides their deployment in embedded use-cases and cyber-physical systems, FPGAs are also deployed in servers and high-performance computing clusters to offer on-demand performance acceleration to complex compute-intensive algorithms~\cite{watanabe2019implementation}.
These reconfigurable platforms include hard IPs (IC realization) of low-power ARM processor cores and commonly used hardware accelerators like video codecs, besides the soft IP cores, which are used to realize the custom accelerators~\cite{crockett2014zynq}.
Due to these wide-ranged capabilities, FPGAs can act as a \textit{Programmable-System-on-chip} (PSoC) for embedded use-cases or as a service for on-demand compute acceleration.
However, FPGAs tend to be more power hungry, with decreased throughput, when compared to ASICs.

Besides the use of conventional power-reduction approaches, like Dynamic Voltage and Frequency Scaling (DVFS)~\cite{mantovani2016fpga} and clock-gating~\cite{huda2009clock}, \textit{Approximate Computing} appears to be quite suitable for increasing the energy-efficiency of a system.
The primary notion behind approximating the computational units of a system, stems from the fact that these systems and their algorithms are inherently error-resilient, \textit{i.e.}, they do not lead to ``significant'' degradation in output quality~\cite{chippa2013analysis}.
This inherent error-resilience property is exhibited by applications encompassing several domains like recognition, mining, and synthesis, to include data analysis~\cite{samadi2014paraprox}, audio, image, or video processing~\cite{el2017embracing}, speech and image recognition~\cite{chippa2014scalable}, etc.
The resilience exhibited by these applications can be attributed to four key factors, namely,
\begin{inlinelist}
    \item redundant input data,
    \item implicit error attenuation by the deployed algorithms,
    \item non-distinguishable, user-level, differences in the output quality, and
    \item lack of a single golden output.
\end{inlinelist}
These factors have been exploited widely in various research works, spanning academia and industry, focusing on both hardware and software to obtain either power, latency, or energy benefits~\cite{saadat2019approximate,hashemi2015drum,ullah2018area,prabakaran2018demas,ullah2022appaxo,prabakaran2020approxfpgas,echavarria2016fau,ullah2021clapped,ullah2021high,ullah2018smapproxlib,vasicek2016search,venkataramani2013quality,mrazek2019autoax}.

The state-of-the-art tends to primarily focus on generating approximate circuits (AC) and/or accelerators to obtain power, latency, and/or energy benefits for ASIC-based systems.
Several past works have illustrated that the approximate computing principles and techniques developed for ASIC-based systems tend to offer dissimilar benefits when realized for FPGA-based systems~\cite{ullah2018area,prabakaran2018demas,ullah2022appaxo,prabakaran2020approxfpgas,echavarria2016fau,ullah2021clapped,ullah2021high,ullah2018smapproxlib,vasicek2016search}.
For instance, approximate arithmetic units developed for ASICs can achieve up to $70\%$ energy reduction when synthesized for ASIC platforms; the same designs offer minimal savings, or at times negative savings, when synthesized for FPGA platforms~\cite{ullah2018area}.
Ullah \textit{et al.}~\cite{ullah2022appaxo} proposed a methodology for systematically generating a wide-range of approximate multiplier architectures that offer similar benefits when deployed on FPGA platforms.
The methodology also includes a small-scale design space exploration stage to identify approximate accelerator designs suitable for the application.
On the other hand, Prabakaran \textit{et al.}~\cite{prabakaran2020approxfpgas} illustrate the benefits of exploring the vast architecture-space of approximate ASIC-based arithmetic circuits to identify a set of Pareto-optimal\footnote{By Pareto-optimal, we mean the best solution obtained by a given method.} designs suitable for the target FPGA platform. 
This is subsequently leveraged by a large-scale automated approximate accelerator exploration strategy to identify near-optimal approximate accelerators for an application.
However, these works do not holistically explore the large architecture-space of approximate accelerators that can be constructed using several arithmetic ACs, which might cause the exploration to overlook certain trade-offs that are clearly situated on the Pareto front for the application. 


\subsection{Motivational Analysis}
\label{subsec:MA}

To further illustrate the differences between ASICs and FPGAs, we explore approximate accelerator variants of a Gaussian filter composed of nine $8$-bit multipliers and eight $16$-bit adders.
The approximate accelerator variants are built using the Pareto-optimal ACs present in the evolutionary approximate arithmetic circuit library~\cite{mrazek2017evoapprox8b} and the ACs identified as Pareto-optimal for FPGAs~\cite{prabakaran2020approxfpgas}.
The range of approximate accelerators obtained by identifying exhaustive combinations of these designs are synthesized using the Vivado $2017.2$ tool-chain for the Xilinx \texttt{xc7vx485tffg1157-1} FPGA.
We also disable the use of DSP blocks on the FPGA to ensure that the accelerator is mapped onto the reconfigurable logic.
The quality of these variants, in terms of average Peak Signal to Noise Ratio (PSNR), are determined using their software models for a set of input signal samples.
These experimental results are illustrated in Fig.~\ref{fig:MA}, from which we make the following \textit{\textbf{key observations}}:
\begin{enumerate}[label=(\arabic*),leftmargin=*]
    \item Approximate accelerators that are Pareto-optimal for ASIC-based platforms are not necessarily Pareto-optimal when synthesized for an FPGA, especially when considering the different hardware parameters, such as power and latency.
    \item The time required for synthesizing a subset of approximate accelerators, composed of $1000$ random designs, is $2.8$ days.
    This overhead can be attributed primarily to the high-level mapping algorithms of the Vivado tool-chain.
    \item Although quite a few research works focus on designing custom-made approximate circuits and accelerators for FPGAs~\cite{prabakaran2018demas,ullah2021high}, they are not as beneficial in achieving similar improvements in hardware costs, when compared to automated generation and exploration of approximate circuits, as reinforced by~\cite{prabakaran2020approxfpgas,mrazek2017evoapprox8b}.
\end{enumerate}

\setcounter{figure}{0}
\begin{figure}[t]
    \centering
    \captionsetup{singlelinecheck=false}
    \includegraphics[width = \linewidth]{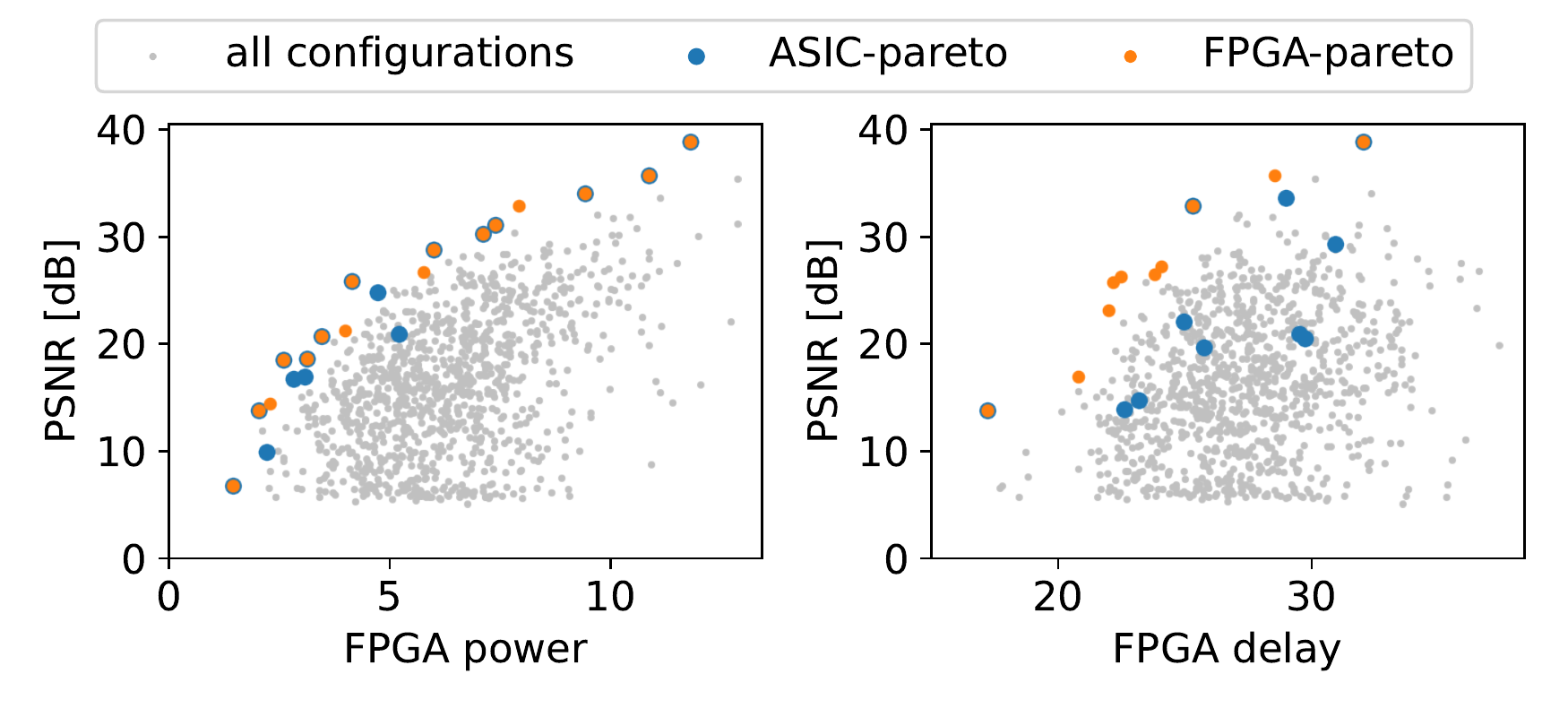}
    \caption{\textbf{Analysis of the Pareto-optimal Approximate Accelerators.}\label{fig:MA}}
\end{figure}

\setcounter{figure}{1}
\begin{figure*}[t]
    \centering
    \captionsetup{singlelinecheck=false}
    \includegraphics[width = \linewidth]{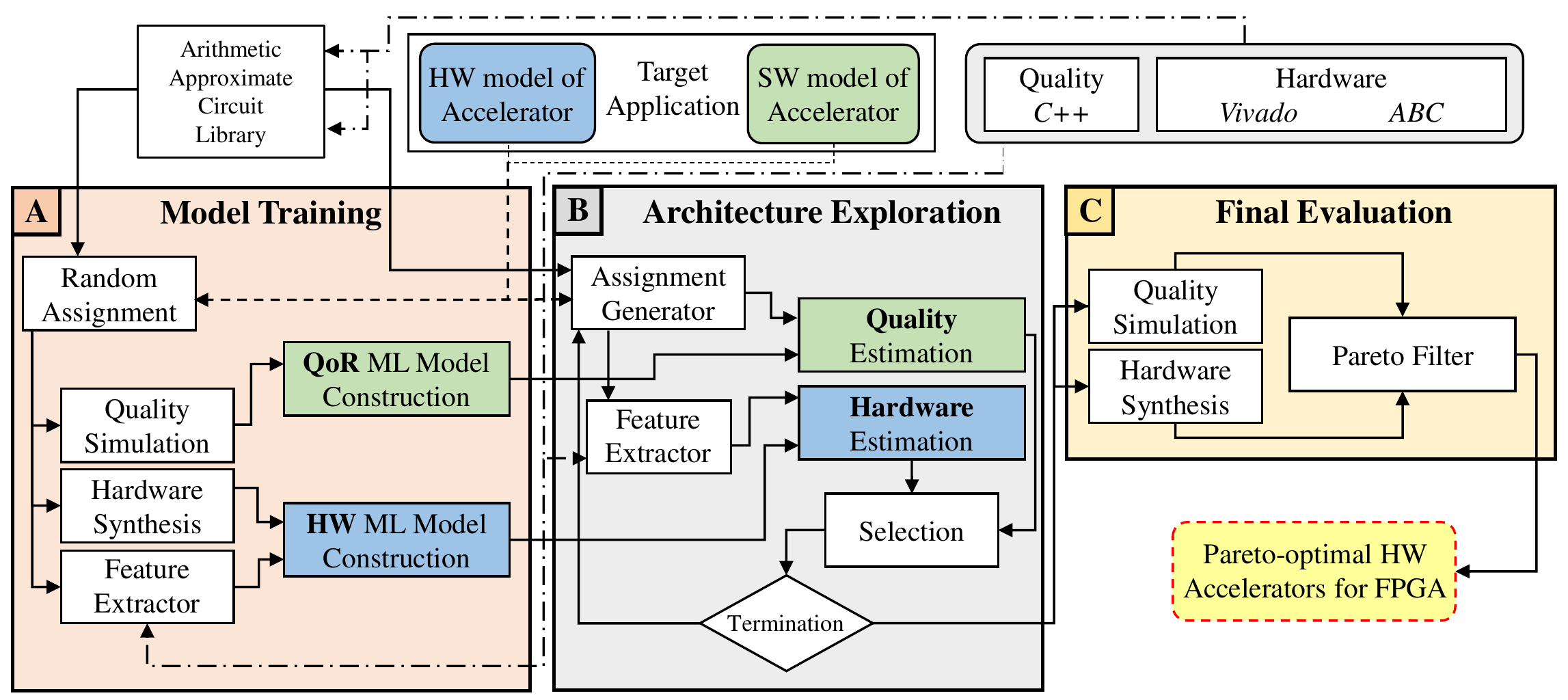}
    \caption{\textbf{An Overview of the Xel-FPGAs Framework.}\label{fig:autoXFPGAs}}
\end{figure*}

From these observations, we have identified the following \textit{\textbf{research challenges}}:
\begin{itemize} [leftmargin=*]
    \item Given the amount of time required to explore the large number of approximate accelerator variants for a simple Gaussian filter, the time required for exploring larger applications would be in the order of months or years.
    \begin{itemize}
        \item \textit{How can we leverage statistical or machine learning models to efficiently reduce the exploration time for FPGA-based approximate accelerators?}
        \item \textit{Can non-conventional features and tools be used, besides the traditional hardware requirements obtained from an FPGA synthesis tool,  to reduce the exploration time?}
    \end{itemize}
    \item There is no single tool that can identify approximate accelerator variants for FPGAs given an application.
    \begin{itemize}
        \item \textit{Can we build an end-to-end framework that can generate and explore approximate accelerator variants for FPGAs given the application's hardware and software models?}
    \end{itemize}
\end{itemize}

To address these research challenges, we propose the following \textbf{\textit{novel contributions}}:
\begin{itemize}[leftmargin=*]
    \item An end-to-end automated framework, \textit{Xel-FPGAs}, for systematically exploring the architecture-space of FPGA-based approximate accelerator variants given a target application.
    \item Our framework employs statistical learning models to traverse the architecture-space of approximate accelerators for FPGAs without synthesizing each individual design.
    \item To further reduce the exploration time of the approximate accelerator architecture-space, we use the ABC tool~\cite{mishchenko2007abc} for estimating the number of lookup tables, power, and latency of the accelerator on our target FPGA.
    \item The complete \textit{Xel-FPGAs} framework, including the approximate accelerators discussed in this work, are open-source and available online at \textcolor{blue}{\url{https://github.com/ehw-fit/xel-fpgas}}, to enable wide-spread adoption and research in this area.
\end{itemize}

\section{The Automated Framework}
\label{sec:autoXFPGAs}

Fig.~\ref{fig:autoXFPGAs} presents an overview of the proposed Xel-FPGAs framework.
The framework is made of three key stages, namely, Model Training, Architecture Exploration, and Final Evaluation.
In the first stage, we train and test various statistical learning models that can be used to estimate the parameters of a given hardware accelerator module.
Next, the models are used to estimate the potential output quality and hardware requirements of all approximate accelerators in the architecture-space.
Finally, the selected accelerator designs are re-evaluated by simulating and synthesizing their software and hardware models, respectively.

\subsection{Model Training}
\label{subsec:ModelTraining}

We start with a set of ACs that can replace the arithmetic units in the target algorithms.
We consider the wide-ranged evolutionary approximate component library~\cite{mrazek2017evoapprox8b}, which is composed of both adders and multipliers, to evaluate the efficacy of our framework.
The use of other adder and multiplier libraries~\cite{ullah2022appaxo} are orthogonal to the use of this library and can be easily incorporated into \textit{Xel-FPGAs}.

The current architecture-space exploration pipeline requires the system designers to generate all possible approximate accelerator variants and synthesize them using an FPGA synthesis tool in order to determine their hardware requirements.
Similar exploration strategies need to be carried out using software models for determining the quality-of-result (QoR) achievable by these variants.
As emphasized earlier, this can lead to an exhaustive architecture-exploration stage that can potentially last for months or even years.
To circumvent this bottleneck, we propose to train and deploy statistical learning models that can be used to estimate the hardware requirements and QoR capabilities of an accelerator variant, thereby drastically reducing the time needed for synthesizing and evaluating each individual approximate accelerator.
Previous works~\cite{prabakaran2020approxfpgas} have evaluated similar models for approximate circuits and determined that \textit{statistical regression} models achieve maximum correlation, in comparison to other approaches like Stochastic Gradient Descent, Adaptive Boosting, or even Multi-Layer Perceptron (MLP).
We focus primarily on \textit{statistical regression} models, which are lightweight, instead of larger machine learning models, to reduce the overall time required for training the models and exploring the architecture-space.

\setcounter{figure}{2}
\begin{figure}[t]
    \centering
    \captionsetup{singlelinecheck=false}
    \includegraphics[width = \linewidth]{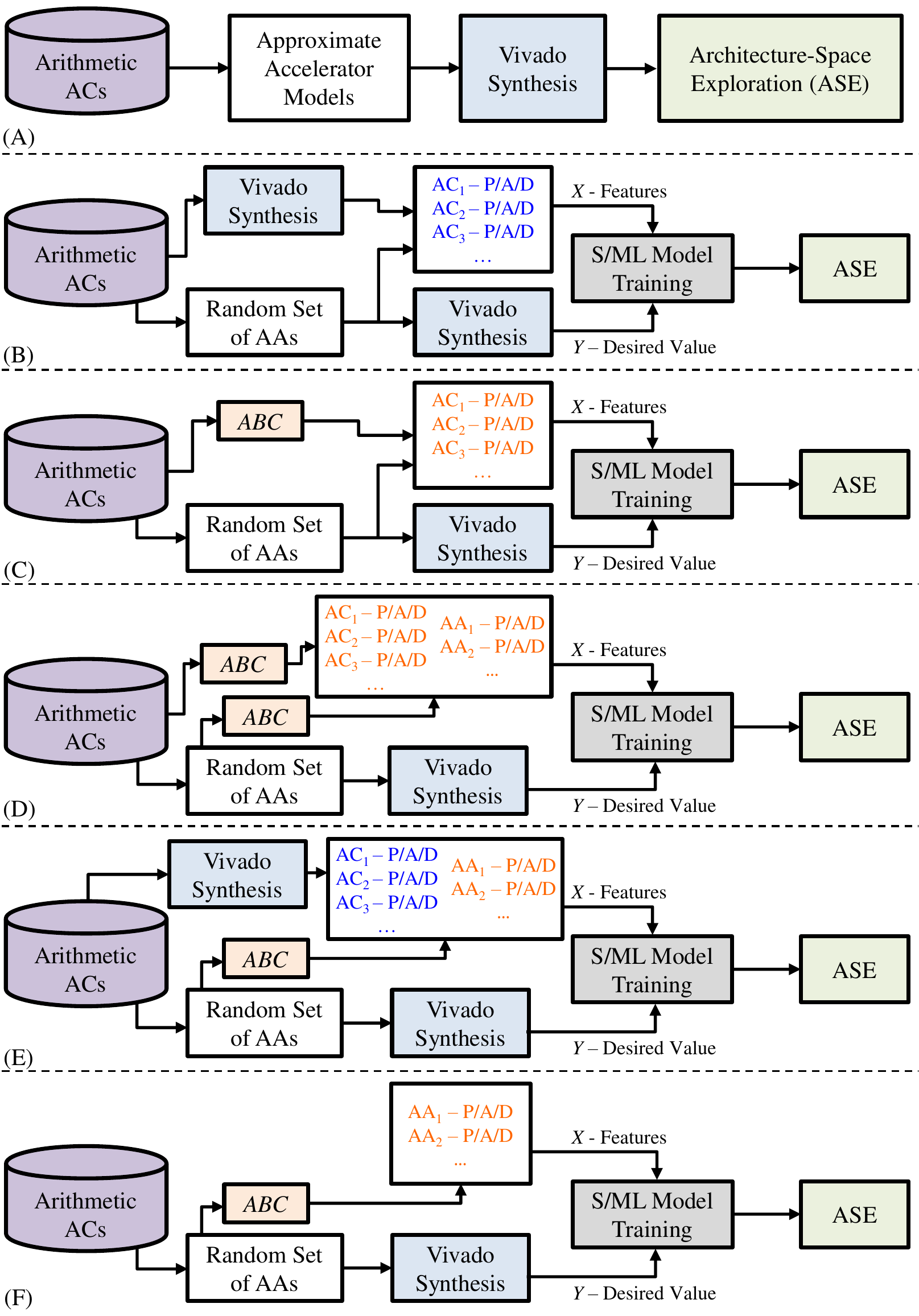}
    \caption{\textbf{Overview of the Possible Model Training Pipelines.}\label{fig:Pipelines}}
\end{figure}

Fig.~\ref{fig:Pipelines} illustrates the model training pipelines explored in our work to determine the optimal setup that can quickly and efficiently explore the architecture-space of approximate accelerators. 
\texttt{(A)} denotes the default exhaustive exploration approach discussed earlier, which requires each approximate accelerator to be synthesized by our FPGA tool-chain.
Since this is not feasible, we propose to generate a feature-set (composed of power, lookup tables, and delay) by synthesizing the arithmetic ACs present in our library.
This feature-set is used to estimate the hardware parameters for a set of $1000$ random approximate accelerators.
These random approximate accelerators need to be subsequently synthesized using the FPGA tool-chain for the model to learn the relation between the features and its desired value.
This pipeline is depicted in \texttt{(B)}.
However, using the Vivado tool-chain for synthesizing the complete library of ACs is a highly time-consuming process (see Section~\ref{sec:Results}).
Therefore, we propose to replace the Vivado tool-chain with the ABC tool~\cite{mishchenko2007abc} to estimate the features necessary for the model to learn.
This pipeline is presented in \texttt{(C)}.
Another variation of this version would be to include more features that could potentially improve the accuracy of our statistical model.
This variant, illustrated in \texttt{(D)}, includes an additional step wherein the approximate accelerator variants' features are extracted by ABC and fed into the model training stage, with the vivado synthesis values of the approximate accelerators.
And finally, to evaluate all possible combinations, we tested out two additional scenarios; one (i) wherein the original feature extraction pipeline, with the Vivado tool-chain, stays in place for the arithmetic ACs, while the ABC tool is used to extract the approximate accelerators' features (presented in \texttt{(E)}), and second (ii) where the arithmetic ACs' features are not estimated and only the approximate accelerators' features, extracted by the ABC tool, are used to train the statistical model (illustrated in \texttt{(F)}).
We have compared  the experimental results of these pipelines and analysed them to determine that configuration \texttt{(D)} is most suitable for our framework (see Section~\ref{sec:Results}).

\subsection{Architecture Exploration}
\label{subsec:AE}

Now that we have the statistical models trained, for estimating the application-level QoR and its hardware requirements, we can efficiently explore the architecture-space of our approximate accelerator variants.
\changed{Despite deploying statistical models to reduce the overhead time, we investigated the use of several state-of-the-art evolutionary algorithms to further decrease the time needed for exploring the architecture-space by selecting specific models that need to be evaluated.}
\changed{For example, we can start with a set of $1000$ random approximate accelerator variants, which are fed to the ABC tool to determine their features.
With this information, we can estimate the approximate accelerators' hardware requirements and its QoR.
The $200$ best, Pareto- and near-optimal, designs are subsequently fed back to the evolutionary algorithms to act as the parent set. 
Using the parent set, we obtain the next generation of $1000$ approximate accelerators that are to be evaluated, from which we select the $200$ best designs that subsequently act as the parent set for the next generation.
This process is repeated until we have $1000$ approximate accelerator variant generations.
We consider the $200$ best approximate accelerator variants obtained from the last generation to be the final set of Pareto-optimal designs, which we have observed to achieve optimal results.}
Further generation and exploration of approximate accelerator variants achieves sub-optimal results or minimalistic improvements when compared to the time needed for evaluating these designs.
Therefore, we terminate the exploration stage and move to the final stage of the framework, where we evaluate its efficacy.

\textbf{Evolutionary Algorithms:}
\changed{
Besides investigating the commonly used NSGA-II~\cite{deb2002fast} genetic algorithm for exploring our design space, we decided to evaluate two other approaches based on local search to investigate their applicability on our design space.
Motivated by the results illustrated in~\cite{mrazek2019autoax}, we chose to implement and evaluate the Hill climber algorithm~\cite{jacobson2004analyzing}, which ultimately turned out to be highly inefficient for exploring our design space.
The algorithm was able to identify only $1$ candidate solution in each iteration stage, which is then fed into our statistical models.
However, these models are built in such a way that they make heavy use of vectorization and parallelism, and the evaluation on a feature-by-feature basis is slow, which is against the ultimate goal of our framework, \textit{i.e.,} to reduce exploration time.
Hence, we chose to explore the Evolution Strategy (ES) $\mu+\lambda$, where $\lambda$ new candidate designs are generated from a set of best individuals ($\mu$).
In contrast to NSGA-II, which depends on ``genetics'' and complex ``crossover'' mechanisms to identify the next generation of designs, ES is only dependent on small random changes in the best individuals.
We iteratively run through each design in a given Pareto-front --under a time constraint-- to obtain new designs, which can be used to extract a new Pareto-frontier.
The designs from this new frontier act as the base for the next generation of designs.}

\changed{Our multi-objective evolutionary algorithm is described in Algorithm~\ref{alg:es}.
The algorithm starts with a set of random parents ($\mu$) using which we generate $\lambda$ new candidates with random mutations on each parent.
The new candidates' power and QoR are parallelly estimated by our statistical models to determine if the design updates the Pareto-frontier $P$.
If a new design is added to $P$, the design becomes a new parent, which can be subsequently used to generate other new candidates.
Otherwise, we iteratively select a new candidate from the existing Pareto-frontier $P$ in order to prevent getting stuck.
}



\begin{algorithm}[t]
\caption{Our Evolutionary Multi-objective Optimization}\label{alg:es}
\begin{algorithmic}\footnotesize

\Input{$RL$  -- set of libraries, $RL=\{RL_1,RL_2,\cdots,RL_n\}$, 
        $M_{HW}$ -- HW costs model, $M_{QoR}$ -- quality model}

\Output{Pareto set $P \subseteq RL_1 \times RL_2 \times \cdots \times RL_n$ }

\Function{EvolutionaryStrategyMOO}{$RL$, $M_{QoR}$, $M_C$}
\State $ParentSet \gets$
\State $\Call{PickRandom}{ RL_1 \times RL_2 \times \cdots \times RL_n}$ \Comment{$\mu$ parents}
\State $P \gets \emptyset$
\While{$\neg TerminationCondition$}
\ForAll{$Parent \in ParentSet$}
    \State $C \gets \{ \Call{Mutate}{Parent}, \cdots \}$ \Comment{$\lambda$ new candidates }
    \State $e_{QoR} \gets M_{QoR}(C)$ \Comment{\footnotesize Estimate the quality of $C_i$}
    \State $e_{HW} \gets M_{HW}(C)$ \Comment{\footnotesize Estimate the HW costs of $C_i$}
    \ForAll{$c_i \in C$}
    \If {\Call{ParetoInsert}{$P, (e_{QoR}, e_{HW}), C$}} 
        \State $Parent \gets C$ \Comment{Replace current parent}
    \EndIf

    \EndFor
    \If{$Parent \notin C$}
        \State $Parent \gets \Call{PickRandomlyFrom}{P}$ \Comment{\footnotesize Prevent stagnation}
    \EndIf
\EndFor
\EndWhile\label{euclidendwhile}
\State \Return{P}
\EndFunction
\end{algorithmic}

\end{algorithm}

\subsection{Final Evaluation}
\label{subsec:Eval}

Once the final set of approximate accelerator variants are obtained, we synthesize and evaluate them to determine their accurate hardware requirements and QoR values, instead of the estimates obtained from the models.
These values are subsequently used to construct a Pareto-frontier, which acts as the final set of Pareto-optimal variants for the target application.
The system designer can then analyze the quality-hardware trade-offs of these designs to deploy the relevant approximate accelerator variant for the application based on system specifications.

\section{Experimental Setup}
\label{sec:ES}

\begin{figure}[t]
    \centering
    \captionsetup{singlelinecheck=false}
    \includegraphics[width = \linewidth]{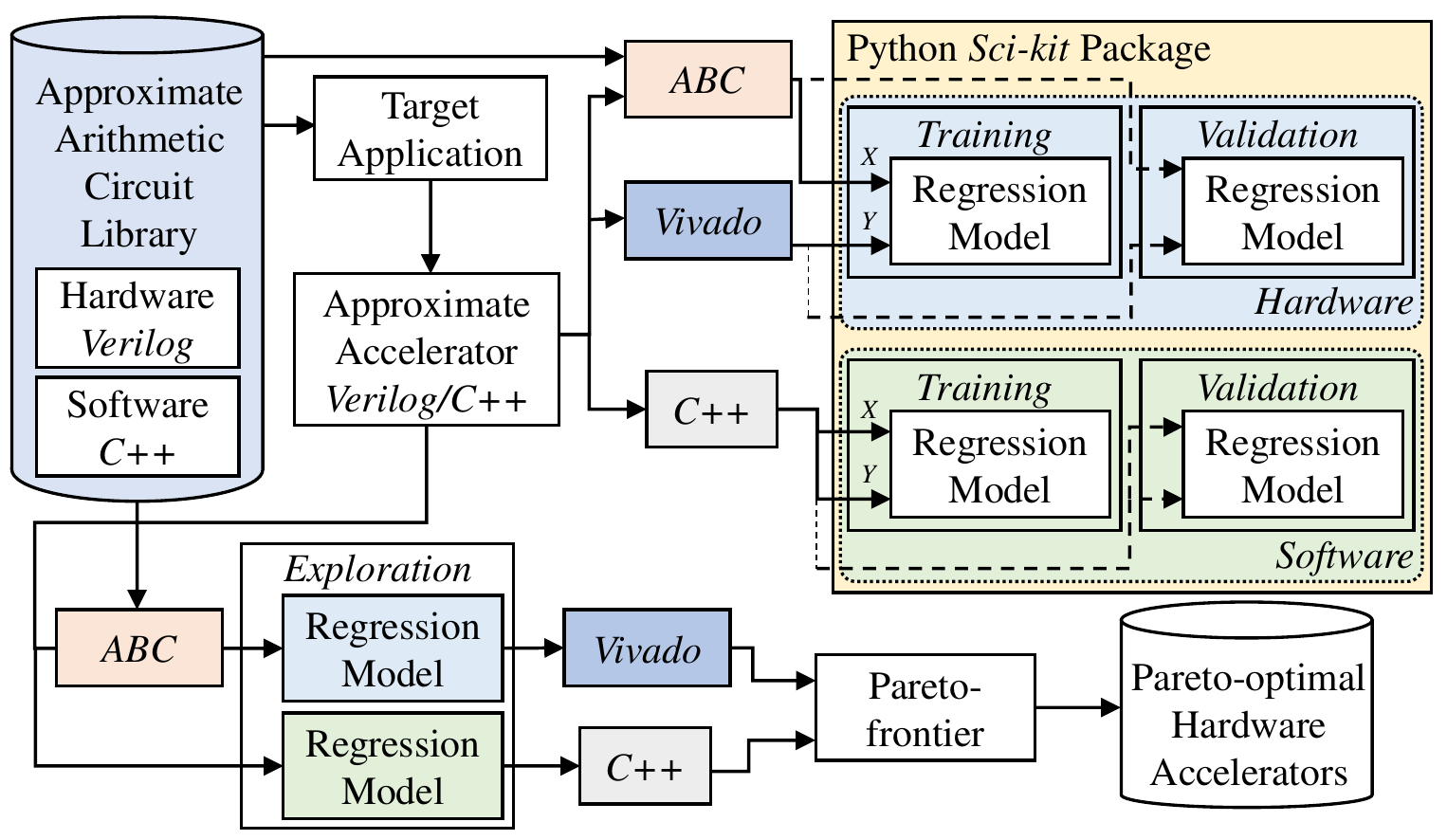}
    \caption{\textbf{An Illustration of our Experimental Setup and Tool-flow.}\label{fig:ES}}
\end{figure}

The approximate arithmetic circuit library used in this work is open-source and immediately accessible\footnote{https://github.com/ehw-fit/evoapproxlib}$^{,}$\footnote{https://github.com/ehw-fit/approx-fpgas}; they include both the hardware (Verilog) and software models (C++) of the ACs.
These can be combinatorially deployed in the target application to generate different possible approximate accelerator variants.
Without loss of generality, in this work we consider a sample set of $1000$ random approximate accelerator variants to train our \textit{statistical regression} model.
Since our framework follows pipeline \texttt{(D)} from Fig.~\ref{fig:Pipelines}, we perform a component- and accelerator-level feature-extraction, from the RTL models, while also synthesizing and mapping the designs, using the Vivado Design Suite 2017.2 for the \texttt{xc7vx485tffg1157-1} Xilinx FPGA, to obtain the relevant hardware parameters, such as power, number of lookup tables, and latency of the approximate accelerator.
We also simulate the application with behavioral models of the accelerator in C++, using the benchmark sequences discussed in~\cite{vasicek2017towards}, to determine the application-level output quality.
Using these values we train two different \textit{regression} models, one for estimating the hardware parameters and the other for estimating the QoR of the approximate accelerator.
The hardware estimator requires the features of the ACs and the approximate accelerators as input to predict the area, power, or latency of the variants, while the software estimator requires the mean and average error of the approximate circuits deployed in the application to estimate its QoR.
Once the models are trained, they are used to selectively and iteratively explore the architecture-space of the accelerator using the NSGA-II and ES evolutionary algorithms, with $\mu$ set to $1$ for the latter.
The scripting for generating the set of approximate accelerator variants, executing the genetic algorithm, and training the statistical learning models are completed in Python.
The final generation of the accelerators obtained this way are synthesized and simulated to determine their accurate hardware requirements and QoR, which are then used to construct the set of Pareto-optimal hardware accelerators for the target application.
An overview of this experimental setup is presented in Fig.~\ref{fig:ES}.

\section{Results and Discussion}
\label{sec:Results}


\begin{figure}[b]
    \centering
    \captionsetup{singlelinecheck=false}
    \includegraphics[width = \linewidth]{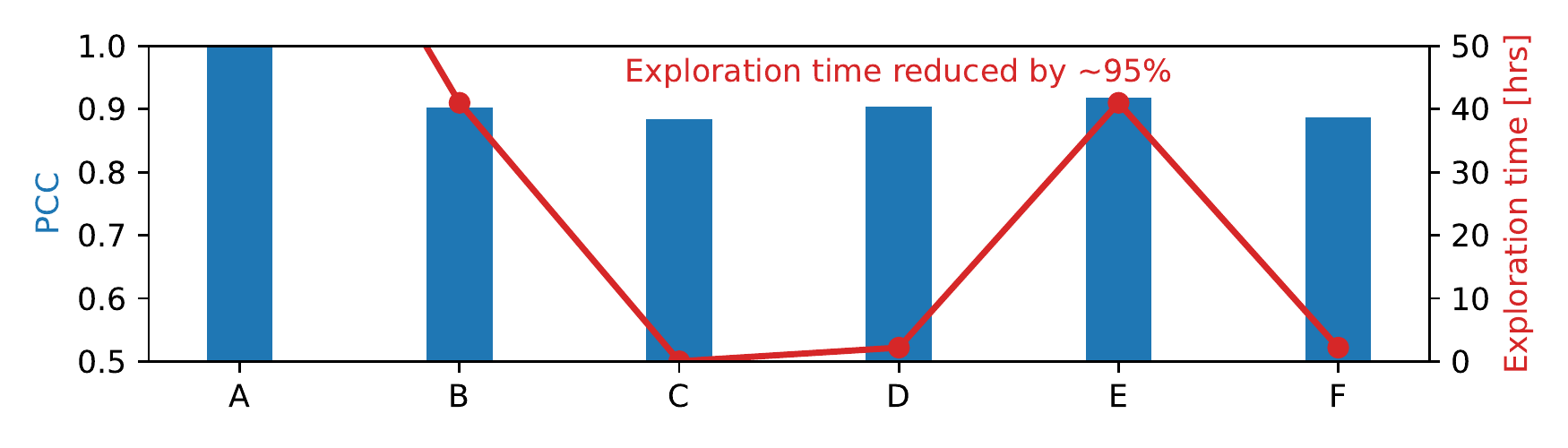}
    \caption{\textbf{Pearson Correlation Coefficient (PCC) and Exploration Time Analysis of the Pipelines Discussed in Fig.~\ref{fig:Pipelines}.}\label{fig:TG}}
\end{figure}

\begin{figure}[t]
    \centering
    \includegraphics[width=\columnwidth]{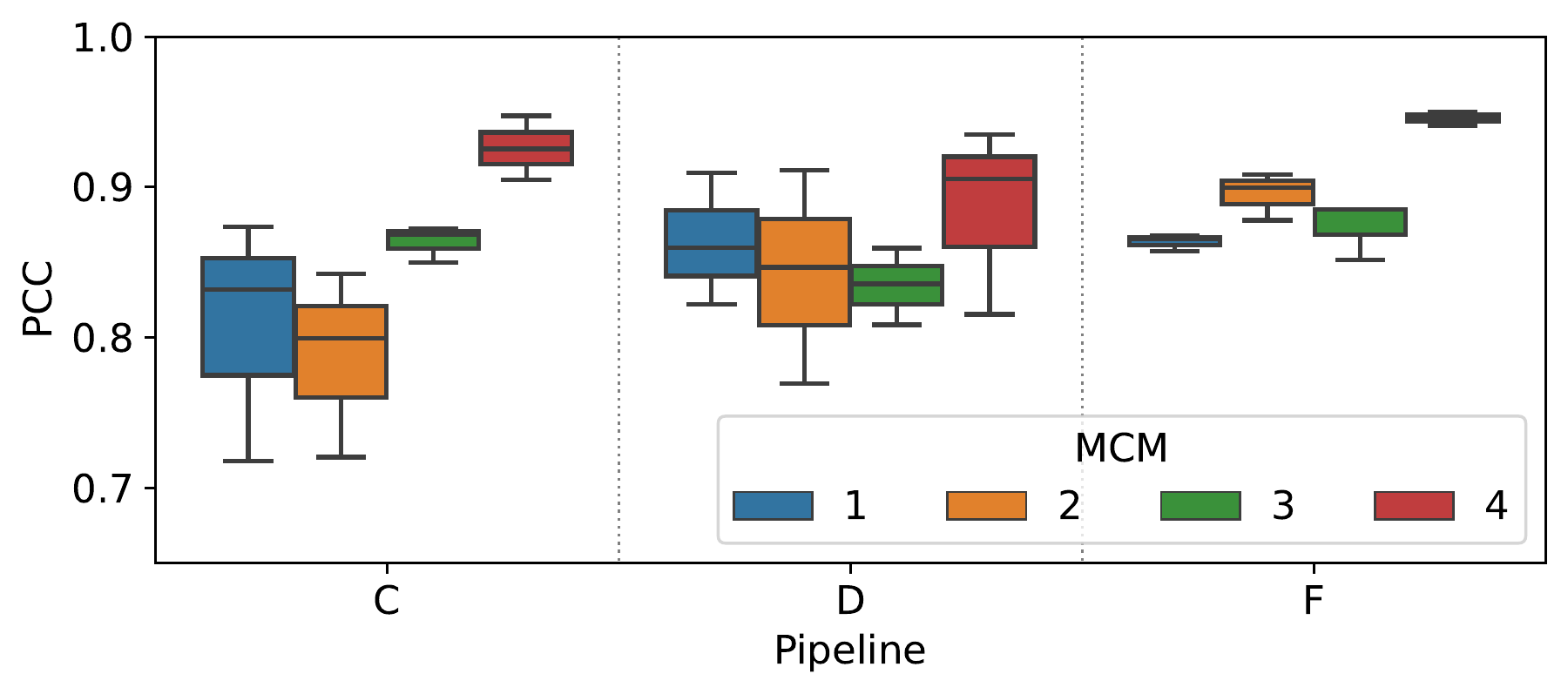}
    \caption{\textbf{Analyzing the PCC of \textit{Xel-FPGAs}' $3$ Fastest Pipelines (\texttt{C}, \texttt{D}, and \texttt{F}) - Estimating the Power of the HEVC Accelerator for Various MCM Modules Deployed in the DCT Block.}}
    \label{fig:pcc}
\end{figure}

We illustrate the efficacy of the \textit{Xel-FPGAs} framework using the High Efficiency Video Coding application.
\changed{HEVC has been shown to be resilient to errors due to its Discrete Cosine Transform (DCT) function.
The DCT block is composed of $8$ to $14$ adders and multipliers, which vary based on the type of Multiplierless Multiple Constant Multiplier (MCM) architecture deployed.
These arithmetic units} can be approximated to increase the system's energy-efficiency and/or performance.
\changed{More information regarding the HEVC accelerator block can be found in~\cite{vasicek2017towards}.}


\subsection{Pipeline Analysis}
First, we evaluate the effectiveness and time requirements of the pipelines illustrated in Fig.~\ref{fig:Pipelines}.
We do this by estimating the correlation between the feature-set established by the pipeline with the desired hardware parameter value (area or power or latency), which is obtained via Vivado Synthesis.
We also evaluate the time required for each pipeline to explore our architecture-space to obtain a set of Pareto-optimal approximate accelerator variants for the target application.
While there is no dependency on the search algorithm for pipelines \texttt{(A)}, \texttt{(B)}, and \texttt{(C)}, the evolutionary algorithm associated selection induces a non-zero evaluation overhead of roughly $30$ms for each approximate accelerator explored using the ABC tool.
Fig.~\ref{fig:TG} illustrates these experimental results.

As anticipated, configuration \texttt{(B)} is successful in reducing the exploration time from thousands of hours (in \texttt{(A)}) to just $41$ hours, which tends to reduce the Pearson correlation coefficient (PCC) to $0.9$.
We can further reduce this to just $2$ minutes by replacing Vivado with ABC in \texttt{(B)}, to obtain \texttt{(C)}, while further dropping the PCC by $2\%$.
Although this may not seem like much, when considering a substantially large architecture-space, such as the HEVC application, the differences in the model estimates can be quite large, leading the algorithm to explore non-optimal designs.
Therefore, an approach that retains the PCC closer to $0.9$ while further reducing the exploration time would be preferable.
Pipeline \texttt{(D)}, which incorporates features from the approximate accelerator variants, achieves a PCC more than $0.9$, while reducing the overhead by nearly $95\%$ for exploring $1,000,000$ approximate accelerator variants.
While configuration \texttt{(E)} achieves a higher PCC, it also increases the exploration time back to $41$ hours, due to the involvement of the vivado tool-chain, whereas with pipeline \texttt{(F)}, the PCC decreases while reducing the exploration time.
Hence, pipeline \texttt{(D)} serves as the ideal configuration for \textit{Xel-FPGAs}.

\changed{
Besides, the reduction in exploration time, we have also performed experiments to analyze the effectiveness of the three fastest pipelines in estimating the power of the HEVC accelerator block when four different types of MCM components are deployed.
The results of these experiments are presented in Fig.~\ref{fig:pcc}.
Although pipeline \texttt{(F)} appears to achieve better solutions, on average, we are interested in the pipeline configuration that obtains the best possible PCC estimates.
For the HEVC use-case, pipeline \texttt{(D)} achieves a trade-off better than the other pipelines. 
}


\subsection{Ablation studies}
Next, we analyze the capability of three different \textit{statistical regression} models, namely, 
\begin{inlinelist}
    \item Random Forest,
    \item Bayesian Regression, and
    \item Support Vector Machines,
\end{inlinelist}
by deploying them in our HEVC use-case.
We explore these models' effectiveness in accurately estimating the QoR and power of the approximate accelerator variants using the PCC metric and pipeline \texttt{(D)}, as established earlier. 
Fig.~\ref{fig:Ablation} illustrates the results of these models for the four different MCM accelerator components deployed in the application.
As can be observed, the different models of \textit{regression} achieve varying levels of correlation based on the parameter being estimated. 
For instance, \textit{Random Forest} achieves the best results when estimating the QoR of all the accelerators in question, \textit{i.e.,} MCM$1$ through MCM$4$.
Whereas \textit{Bayesian Regression} achieves the best power estimates for all the accelerators.
Therefore, we use these two models, of the nearly $20$ models pre-available in the \textit{Xel-FPGAs} framework, to estimate the relevant parameters when exploring the architecture-space.

\setcounter{figure}{6}
\begin{figure}[t]
    \centering
    \captionsetup{singlelinecheck=false}
    \includegraphics[width = \linewidth]{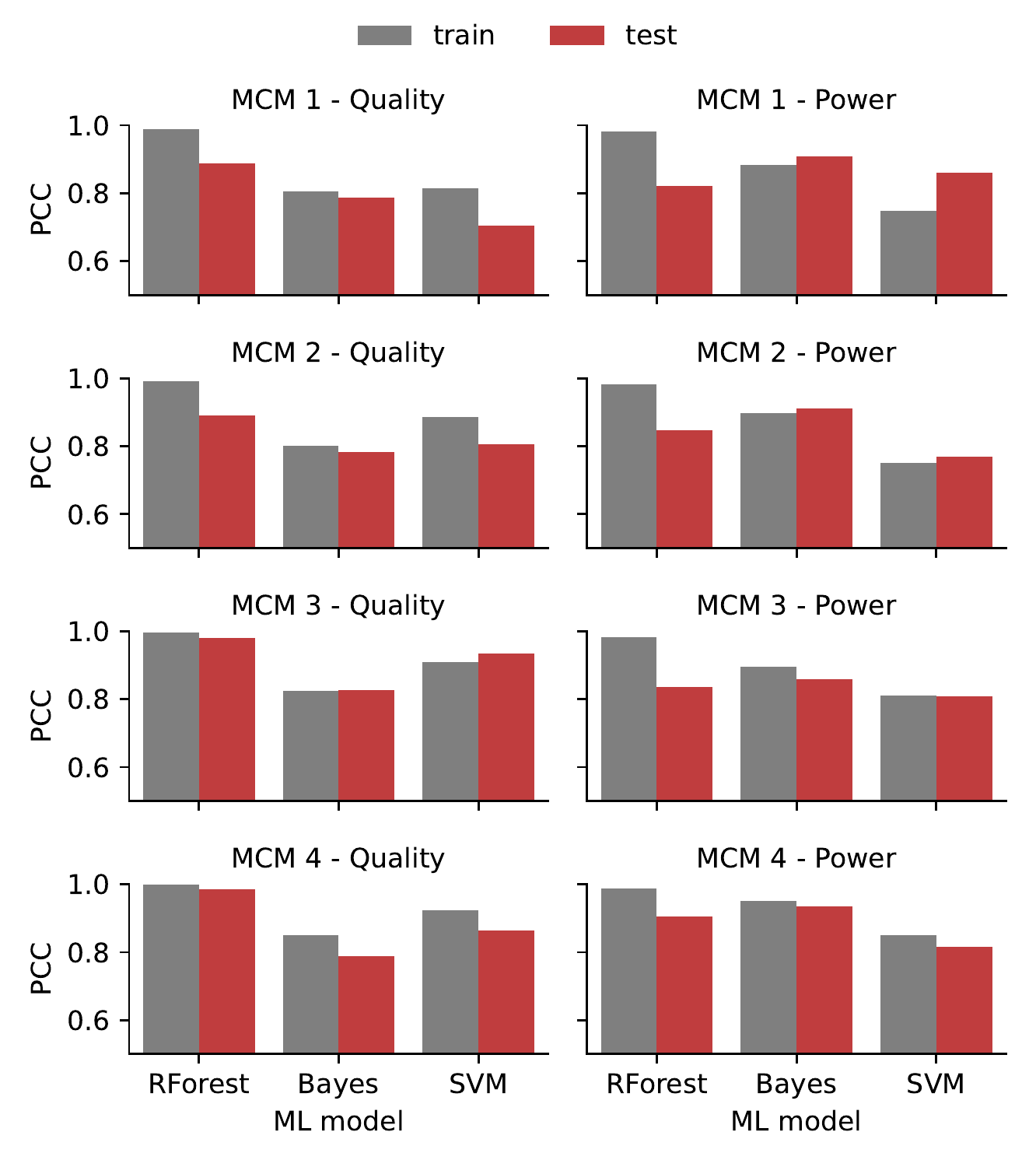}
    \caption{\textbf{Pearson Correlation Coefficient (PCC) Analysis for Three Different Statistical Regression Models that are Used to Estimate HEVC Accelerators' Hardware and Quality Parameters.}\label{fig:Ablation}}
\end{figure}


\subsection{Architecture Exploration}



\begin{figure}[t]
    \centering
    \includegraphics[width=\columnwidth]{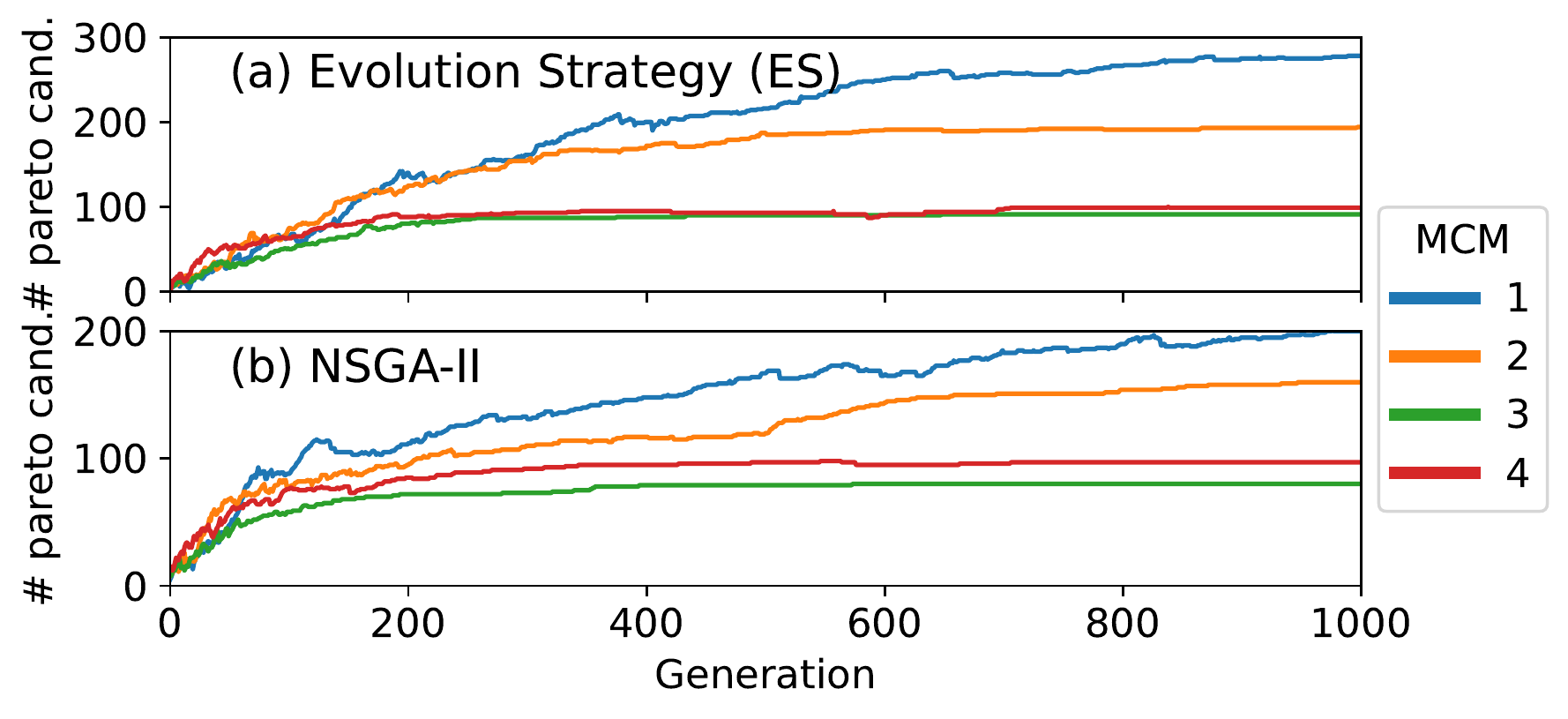}
    \caption{\textbf{Analysing the Convergence of ES and NSGA-II.}}
    \label{fig:es:analysis}
\end{figure}

\setcounter{figure}{9}
\begin{figure*}[b]
    \centering
    \captionsetup{singlelinecheck=false}
    \includegraphics[width = \linewidth]{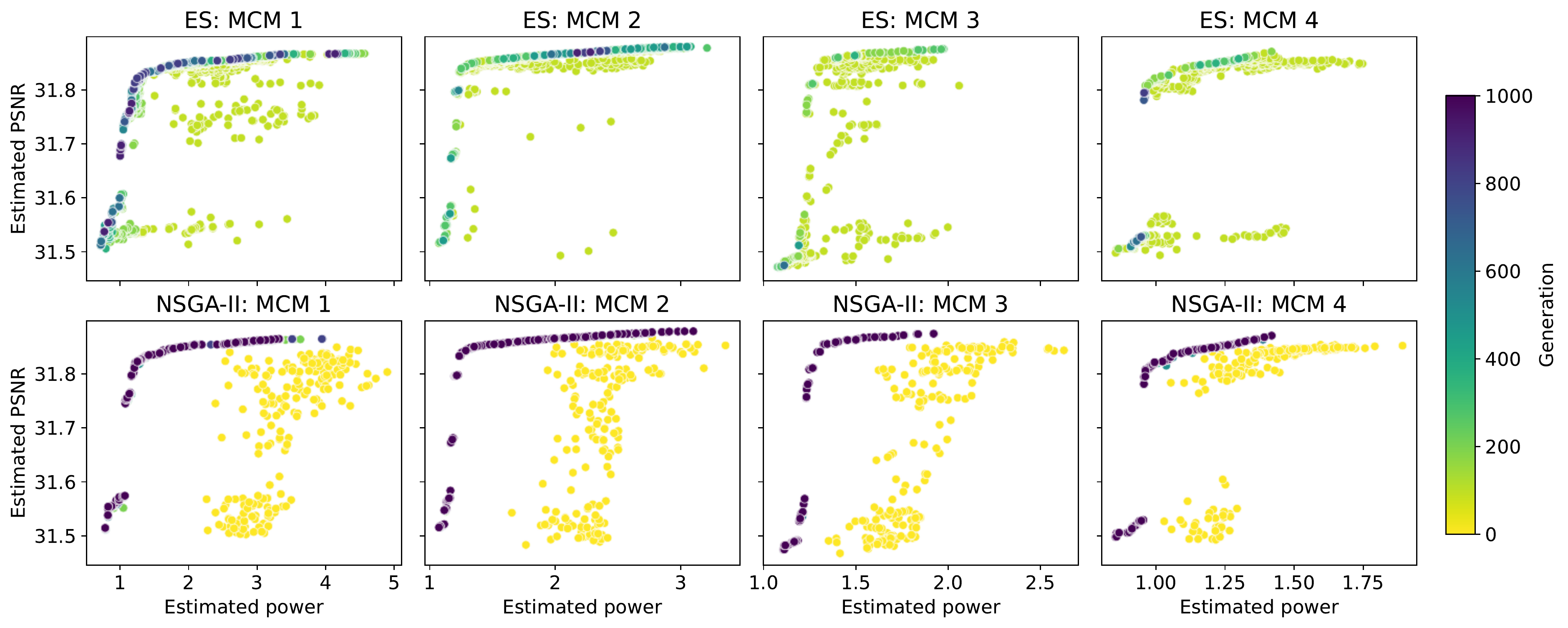}
    \caption{\textbf{Analysis of the Pareto-optimal HEVC Approximate Accelerators from Each Generation of ES (top) and NSGA-II (bottom).}\label{fig:Gen}}
\end{figure*}

\setcounter{figure}{8}
\begin{figure}[t]
    \centering
    \includegraphics[width=\columnwidth]{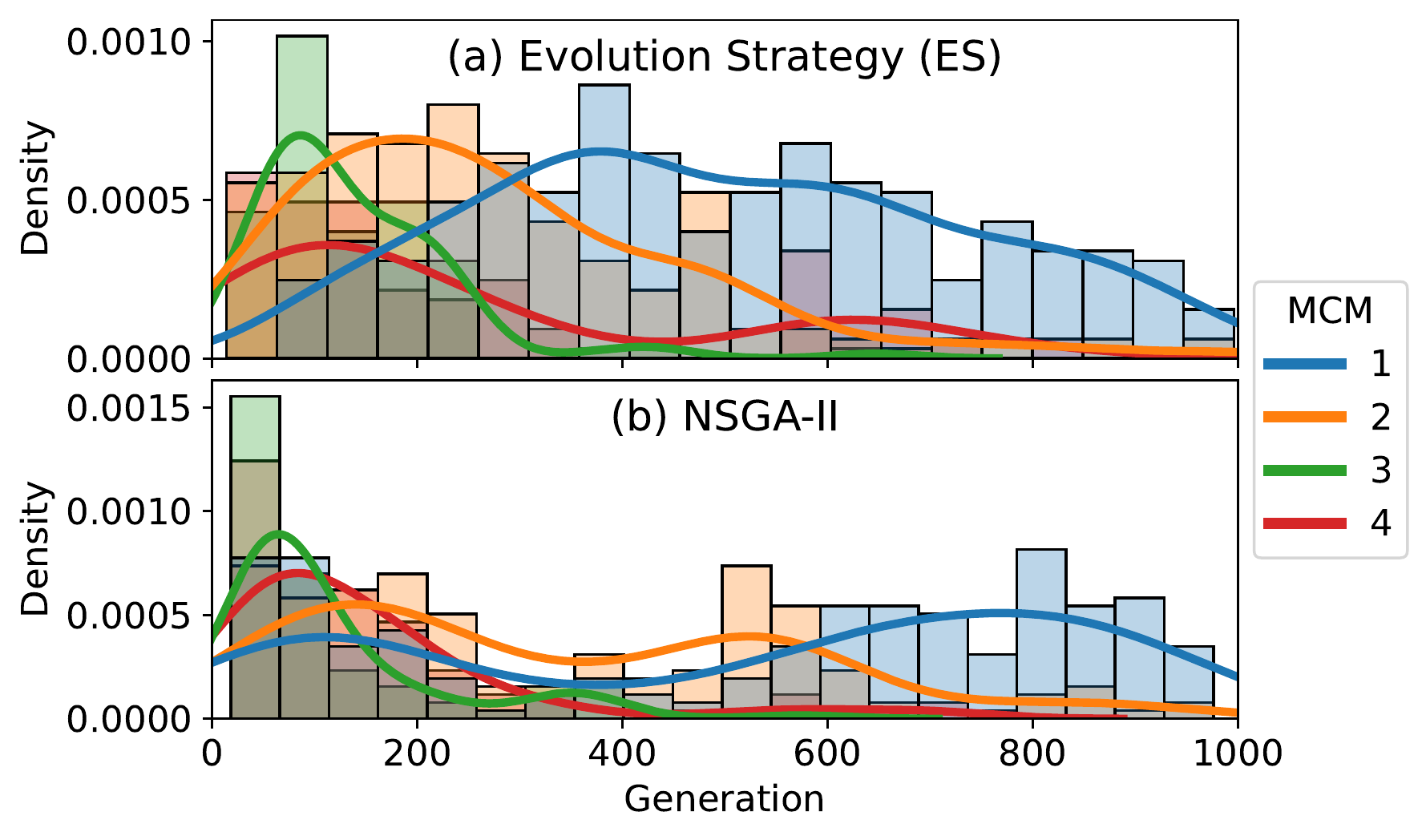}
    \caption{\textbf{Analyzing the Percentage of Total Pareto-optimal Designs (\textit{Density}) Generated by ES and NSGA-II in Each Generation.}}
    \label{fig:es:hist}
\end{figure}

\changed{
Since we have determined the optimal regression models and their corresponding configurations for the use-case with \textit{Xel-FPGAs}, we now move on to analyzing the framework's model generation and exploration capabilities.
Our assumption is that each new generation of the models explored tends to improve the trade-offs, with respect to the previous generation, thereby leading to a set of Pareto-optimal hardware accelerators by the end of our search.
We do this by first analyzing the capabilities of the evolutionary algorithms used in our framework.
Besides the Hill climber algorithm, which was not very beneficial for our use-case, we have evaluated the NSGA-II and the Evolution Strategy (ES) algorithms on the HEVC application, as stated in Section~\ref{subsec:AE}.
These algorithms were executed with similar population settings and ran with the same design time overhead.
Fig.~\ref{fig:es:analysis} illustrates the number of Pareto-optimal points obtained by the ES and NSGA-II algorithms in each generation of their execution. 
Fig.~\ref{fig:es:hist}, on the other hand, illustrates the percentage of total number of Pareto-optimal designs obtained, termed \textit{density}, in each iteration of algorithm execution.
From the former, we can observe that the simpler MCM$3$ and MCM$4$ modules converge faster, \textit{i.e.,} reach a maximum earlier in the number of Pareto-optimal designs obtained, as opposed to the more complex MCM$2$ and MCM$1$ modules.
The simpler modules have hardly achieved better solutions since generation $300$, which is more clearly visible in ES.
On the other hand, NSGA-II requires a few more generations of search before achieving a similar number of Pareto-optimal designs when compared to ES.
The reason behind this becomes clearer when we analyze Fig.~\ref{fig:es:hist}.
A large portion of the Pareto-optimal designs obtained by the NSGA-II algorithm are obtained in the first $300$ iterations, especially for the more complex modules. 
Hence, a second peak is observed for the MCM$1$ and MCM$2$ modules where-in another large percentage of the Pareto-optimal designs are identified.
Since the NSGA-II algorithm starts with a large population set it is able to identify a significant percentage of the Pareto-optimal designs in the earlier generation of search.
The same is not true for the ES algorithm, which has a more uniform distribution in identifying the Pareto-optimal designs through the search iterations. 
Hence, ES might be more suitable for identifying a large number of design trade-offs for the target application, whereas NSGA-II might be better suited to achieve good quality solutions quickly, which contradicts our initial assumption.}

\changed{Next, we move on to the main goals of our framework. 
Fig.~\ref{fig:Gen} illustrates the result of generating hardware accelerators using the ES and NSGA-II algorithms, wherein we depict the estimated PSNR and power values of the approximate accelerator variants searched by each generation.}
Although the points identified by \textit{Generation} $0$ illustrate a random distribution, as expected, only \textit{Generation} $1000$'s outcomes are clearly visible while the others outcomes appear to be missing.
This is because the outcomes of \textit{Generation} $1000$ are masking the outcomes of the other generations, \textit{i.e.}, the newer generation outcomes mask the earlier generation outcomes, because their outcomes are similar to the past generations. 
This offers the \textit{opportunity of further reducing the exploration time by a factor of $10$}, to reduce the overall exploration time by $99\%$.
\changed{Moreover, as anticipated, there are a lot more Pareto-optimal points that are obtained by the ES algorithm when compared to NSGA-II search.}






\subsection{Final Evaluation}

\setcounter{figure}{10}
\begin{figure}[t]
    \centering
    \captionsetup{singlelinecheck=false}
    \includegraphics[width = \linewidth]{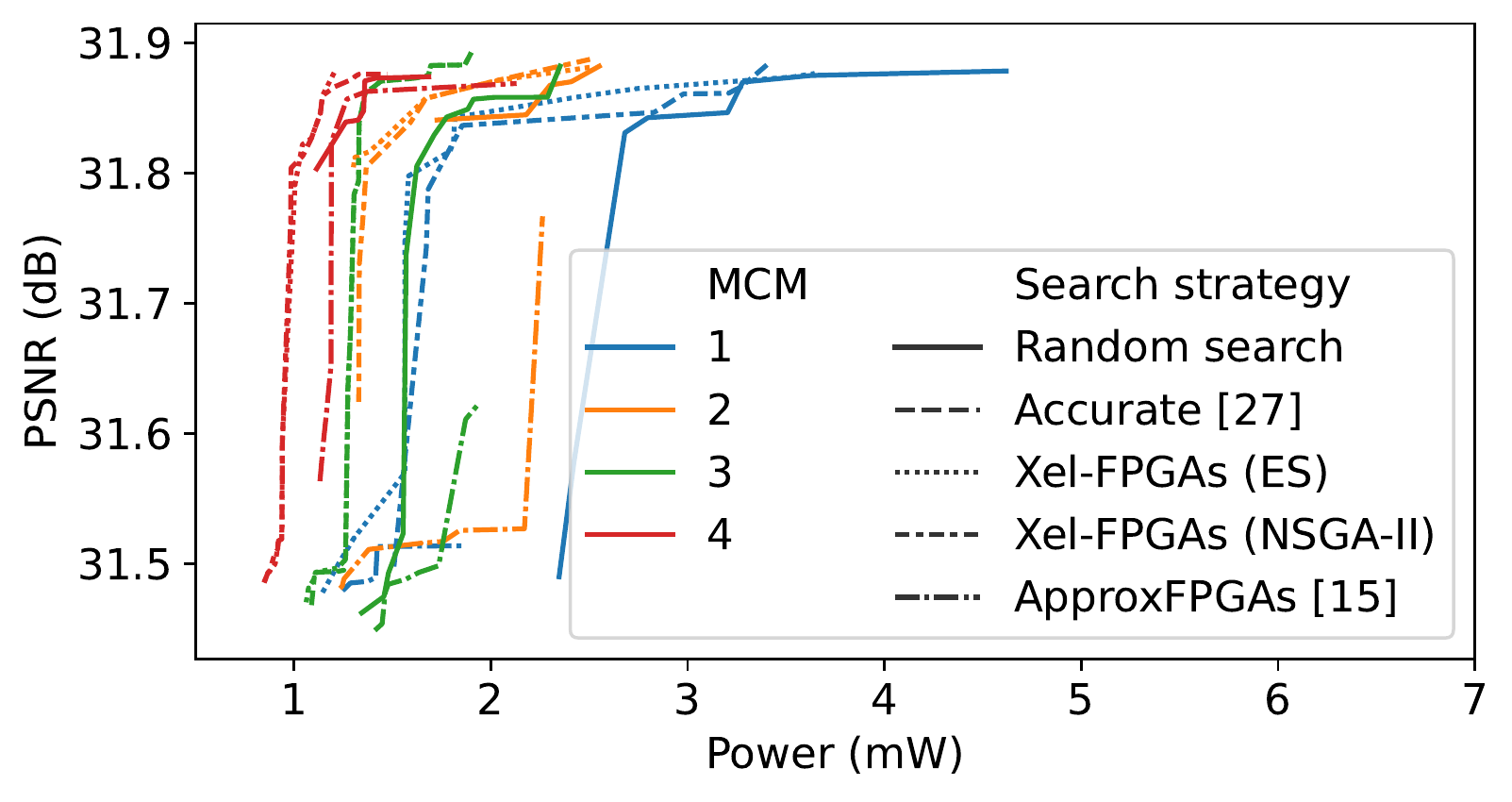}
    \caption{\textbf{Comparison of the Approximate Accelerators Obtained by our \textit{Xel-FPGAs} Framework and the State-of-the-Art~\cite{prabakaran2020approxfpgas}.}\label{fig:MCM}}
\end{figure}

\begin{figure}[t]
    \centering
    \captionsetup{singlelinecheck=false}
    \includegraphics[width = \linewidth]{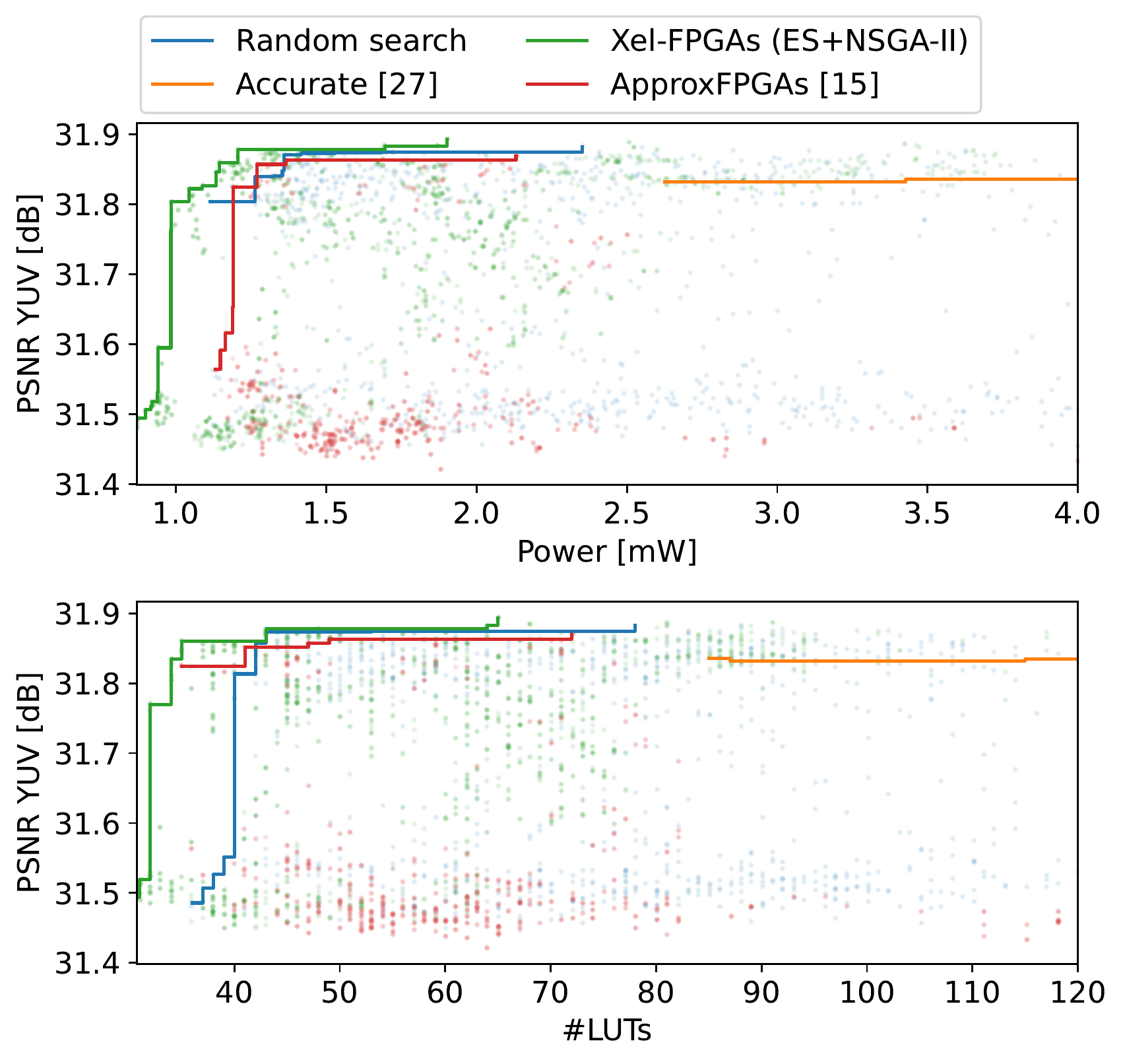} 
    \caption{\textbf{Final Comparison Between \textit{Xel-FPGAs} and the \textit{SoA}~\cite{prabakaran2020approxfpgas}.}\label{fig:Final}}
\end{figure}

With all the intermediate evaluations complete, we move on to to evaluating the overall effectiveness of the framework and comparing it with the current state-of-the-art in approximate accelerator exploration.
We have re-implemented the technique discussed in ApproxFPGAs~\cite{prabakaran2020approxfpgas}, to generate approximate accelerator variants with a set of Pareto-optimal FPGA-based ACs for our HEVC use-case.
Fig.~\ref{fig:MCM} illustrates the accelerator-level comparison between the set of Pareto-optimal approximate accelerator variants obtained by our \textit{Xel-FPGAs} framework and the state-of-the-art ApproxFPGAs approach.
As can be observed, for each of the four accelerator types (MCM$1$ through MCM$4$), our framework outperforms the state-of-the-art and random search to obtain a better set of Pareto-optimal approximate accelerator variants for the HEVC application.
\changed{The use of ES increases the number of Pareto-optimal designs obtained, \textit{i.e.,} it further pushes the Pareto-frontier by obtaining candidate designs that lie on the Pareto-frontier.
Similarly, the combination of evolutionary algorithms in the \textit{Xel-FPGAs} framework clearly} identifies better design trade-offs in comparison to ApproxFPGAs, when we perform an application-level analysis of the Pareto-optimal approximate accelerators obtained, as illustrated by the Pareto-fronts in Fig.~\ref{fig:Final}.
\changed{Moreover, we have also managed to identify design points that reduce the area requirements of the applications, in terms of the number of lookup tables required, which is another benefit of using our \textit{Xel-FPGAs} framework.}




\subsection{Scalability}
\label{subsec:Scale}
\begin{figure}[t]
    \includegraphics[width=\columnwidth]{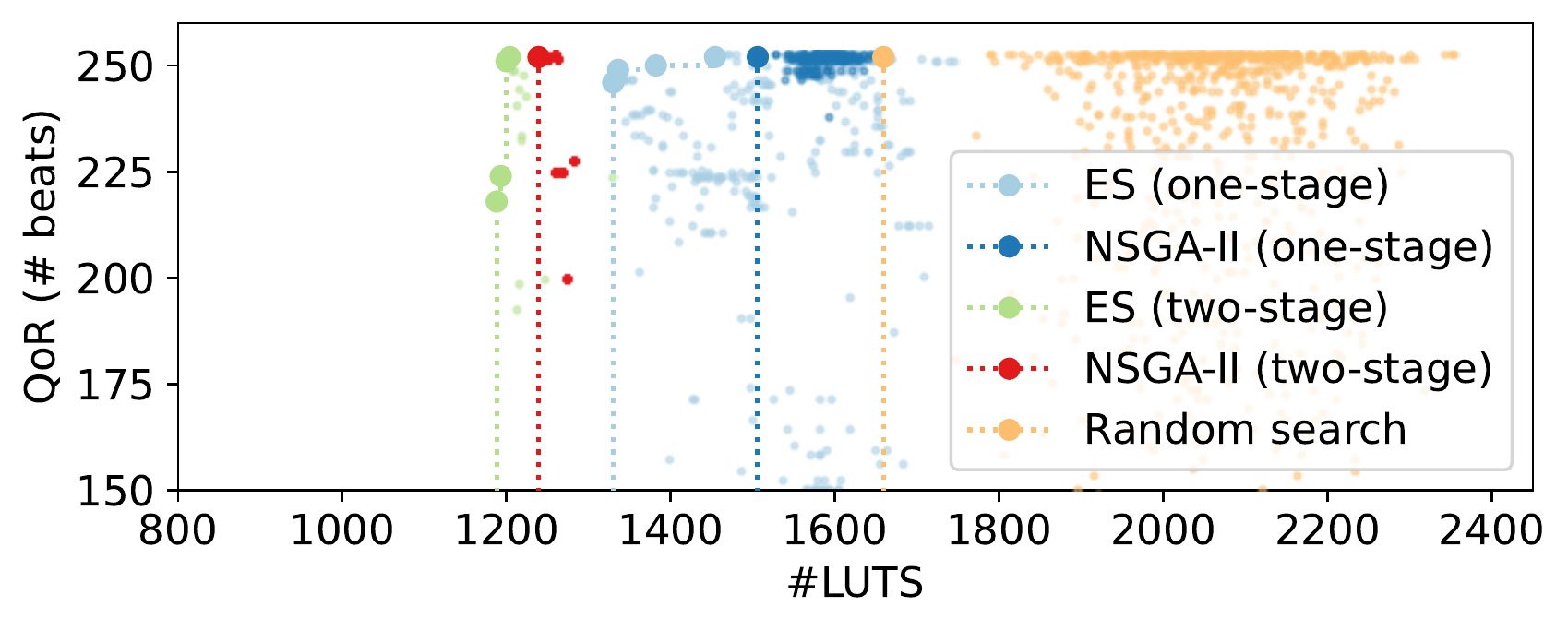}
    \caption{\textbf{Analyzing the Scalability of \textit{Xel-FPGAs}.}}\label{fig:multistage}
\end{figure}

\changed{
Finally, in order to evaluate the ability of our framework to scale for a multi-stage application, we consider an approximate version of the bio-signal processing Pan-Tompkins algorithm discussed in~\cite{prabakaran2019xbiosip}.
The algorithm is made up of five application stages each of which is a filter composed using a varying combination of adders and multipliers, accumulating to $73$ arithmetic units.
Hence, the primary goal of this experiment is to evaluate whether the framework is better suited for decomposing the exploration into multiple-stages (\textit{n-stage}), \textit{i.e.,} a search for each application stage followed by a combined search with Pareto-optimal designs from each stage, or to design everything at once (\textit{$1$-stage}).
As emphasized in the article~\cite{prabakaran2019xbiosip}, we use the application-level quality metric --number of heartbeats-- to evaluate the quality of an application stage by embedding the behavioral model of an approximate variant into the application.
Fig.~\ref{fig:multistage} presents an overview of the results obtained when using the \textit{$1$-stage} and \textit{$2$-stage} (in this case, $n$$=$$2$) exploration strategy.
We can observe that the \textit{$2$-stage} strategy designs, obtained from both ES and NSGA-II, significantly out-perform the designs obtained from the \textit{$1$-stage}.
Hence, we can conclude that the hierarchical and successive approximation strategy (\textit{$n$-stage}) is better at identifying an improved set of Pareto-optimal designs, which are more suitable for the application.
}


\color{black}
\section{Conclusion}
\label{sec:Conc}

We presented \textit{Xel-FPGAs}, which is an end-to-end automated framework for architecture-space exploration of approximate accelerators given a target application that is to be deployed in FPGAs.
We train and use \textit{statistical regression} models to estimate the hardware requirements and output quality of an application, thereby effectively reducing the time needed for exploring the architecture-space.
We also propose to use the ABC tool as a feature extractor for the ACs and our approximate accelerators, to further reduce the  exploration time of our framework while not lowering the quality of our estimations.
Using these features we efficiently explore the architecture-space of approximate accelerators, iteratively using the NSGA-II algorithm, to identify a set of $200$ Pareto-optimal approximate accelerators at the end of $1000$ generations.
These designs are re-synthesized and simulated to compute their accurate hardware requirements and quality to determine the final set of Pareto-optimal approximate accelerator variants for the target application.
The trade-offs between these approximate accelerator variants can be analyzed by the system designer to select one that meets the system specification and quality requirements of the application.
The \textit{Xel-FPGAs} framework outperforms the state-of-the-art in identifying a better set of Pareto-optimal approximate accelerator variants, while also reducing the exploration time by nearly $95\%$.
\changed{We have also illustrated the scalability of our framework using a multi-stage application to discuss the benefits of using a hierarchical search strategy.}
Our \textit{Xel-FPGAs} framework is open-source and accessible online at \textcolor{blue}{\url{https://github.com/ehw-fit/xel-fpgas}}.

\section*{Acknowledgment}
This work was partially supported by the Doctoral College Resilient Embedded Systems, which is run jointly by the TU Wien's Faculty of Informatics and the UAS Technikum Wien, and partially by the Czech Science Foundation project 21-13001S.
This research is also supported by ASPIRE, the technology program management pillar of Abu Dhabi’s Advanced Technology Research Council (ATRC), via the ASPIRE Awards for Research Excellence.

\bibliographystyle{IEEEtran}
\bibliography{References}

\end{document}